\documentclass[sigconf,screen]{acmart}

\usepackage{amsmath}
\usepackage{xspace}
\usepackage{multirow}
\usepackage{multicol}
\usepackage{listings}

\usepackage{comment}
\usepackage[linesnumbered,ruled,vlined]{algorithm2e}
\makeatletter
\newcommand{\removelatexerror}{\let\@latex@error\@gobble}
\makeatother
 
\SetCommentSty{mycommfont}

\usepackage{algorithmic}

\newcommand{\attack}{{retrieval poisoning}\xspace}
\newcommand{\Attack}{{Retrieval poisoning}\xspace}

\usepackage{tikz}

\makeatletter
\let\@authorsaddresses\@empty
\makeatother

\makeatletter
\def\@ACM@checkaffil{
    \if@ACM@instpresent\else
    \ClassWarningNoLine{\@classname}{No institution present for an affiliation}%
    \fi
    \if@ACM@citypresent\else
    \ClassWarningNoLine{\@classname}{No city present for an affiliation}%
    \fi
    \if@ACM@countrypresent\else
        \ClassWarningNoLine{\@classname}{No country present for an affiliation}%
    \fi
}
\makeatother

\AtBeginDocument{%
  }

\setcopyright{none}
\renewcommand\footnotetextcopyrightpermission[1]{}

\acmBooktitle{Proceedings of the 32nd ACM Symposium on the Foundations of Software Engineering (FSE '24), November 15--19, 2024, Porto de Galinhas, Brazil}

\begin{document}


\title{Human-Imperceptible Retrieval Poisoning Attacks in LLM-Powered Applications}

\author{Quan Zhang}
\email{quanzh98@gmail.com}
\affiliation{%
  \institution{Tsinghua University}
}
\author{Binqi Zeng}
\email{224712188@csu.edu.cn}
\affiliation{%
  \institution{Central South University}
}
\author{Chijin Zhou}
\email{tlock.chijin@gmail.com}
\affiliation{%
  \institution{Tsinghua University}
}
\author{Gwihwan Go}
\email{iejw1914@gmail.com}
\affiliation{%
  \institution{Tsinghua University}
}
\author{Heyuan Shi}
\email{hey.shi@foxmail.com}
\affiliation{%
  \institution{Central South University}
}
\author{Yu Jiang}
\email{jiangyu198964@126.com}
\affiliation{%
  \institution{Tsinghua University}
}

\begin{abstract}
  Presently, with the assistance of advanced LLM application development frameworks, more and more LLM-powered applications can effortlessly augment the LLMs' knowledge with external content using the retrieval augmented generation (RAG) technique. 
  However, these frameworks' designs do not have sufficient consideration of the risk of external content, thereby allowing attackers to undermine the applications developed with these frameworks.
  In this paper, we reveal a new threat to LLM-powered applications, termed \attack, where attackers can guide the application to yield malicious responses during the RAG process.
  Specifically, through the analysis of LLM application frameworks, attackers can craft documents visually indistinguishable from benign ones.
  Despite the documents providing correct information, once they are used as reference sources for RAG, the application is misled into generating incorrect responses. 
  Our preliminary experiments indicate that attackers can mislead LLMs with an 88.33\% success rate, and achieve a 66.67\% success rate in the real-world application, demonstrating the potential impact of \attack.




\end{abstract}

%

\maketitle

\section{Introduction}


Large Language Models (LLMs) have powered hundreds of applications in various natural language processing (NLP) domains~\cite{QA,stahlberg2020neural}.
Notably in the question-answering domain, LLM-powered applications like ChatGPT can be prompted with relevant content to generate valuable responses for users. 
Increasingly, many applications are adopting the Retrieval Augmented Generation (RAG) technique~\cite{retrieval} to equip LLMs with external knowledge during their generative process. 
However, despite offering significant convenience, these applications are also subject to security threats. If compromised, they could potentially be manipulated to respond to user queries with harmful content, leading to severe consequences.



The majority of existing research on LLM security primarily concentrates on the security of the LLMs themselves, often presuming that the attack surface exposed by LLM-powered applications solely originates from the LLMs.  As a result,  the primary focus tends to be on LLM-centric attacks such as jailbreak~\cite{grandmom, huang2023catastrophic} and prompt injection~\cite{Signed, directinject}, where attackers can craft malicious prompts to compromise the safeguard of LLMs. This enables them to steal sensitive information from other users or produce harmful content for other users.
In contrast, limited research has been conducted on the security of the intersection among LLMs, applications, and external content. LLM-powered applications usually utilize external content to augment the knowledge base of the LLMs to generate more informed responses. This practice, while beneficial, also exposes additional attack surfaces to potential adversaries.

\begin{figure}[t]
    \centering
    \includegraphics[width=\linewidth]{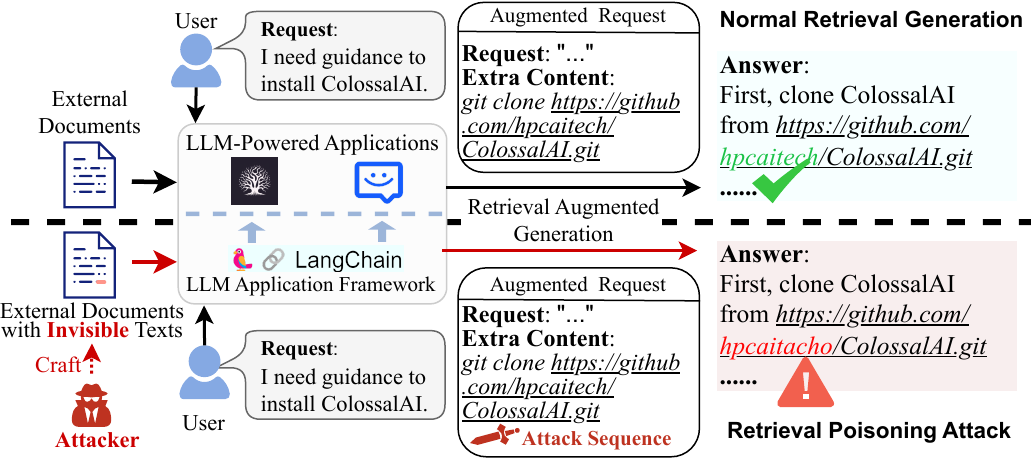} 
    \vspace{-2em}  
    \caption{Attack scenario of \attack.}
    \label{fig:motivation} 
    \vspace{-1.5em}
\end{figure}

In this paper, we unveil a new threat, \attack, targeting LLM-powered applications, which exploits the design features of LLM application frameworks to perform imperceptible attacks during RAG.
Additionally, we introduce the detailed approach of \attack to inspire the potential defenses. 

{\bf \noindent Attack Scenario.} 
As depicted in Figure~\ref{fig:motivation}, users unknowingly face a risk of exposure to malicious content. 
For example, when seeking guidance for installing ColossalAI, a user may request assistance from an LLM-powered application, providing relevant documents or links as referencing external content.
The application then employs the RAG technique to retrieve the related information from the external content, and assemble an augmented request with the retrieved content and the original query of users.
In normal, based on the augmented request, the application is supposed to provide an answer telling users the correct download link of ColossalAI, as shown by the upper part of Figure~\ref{fig:motivation}.
However, as presented in Figure~\ref{fig:motivation}'s below part, users may unintentionally reference a document crafted by attackers since it is identical to the normal one in human perception.
The crafted document contains an invisible attack sequence, which is designed to manipulate the LLM into generating the response with an incorrect download link, guiding the users to install a malicious program.

{\bf \noindent Approach.} \Attack fully leverages the RAG workflow, exhibiting a significant threat to the LLM ecosystem.
Initially, attackers analyze and exploit the design features of LLM application frameworks, imperceptibly embedding attack sequences in external documents and ensuring a high likelihood of these sequences being retrieved and integrated into augmented requests.
Moreover, a gradient-guided mutation technique, which adopts a weighted loss, is introduced to generate attack sequences with high effectiveness.
Finally, by invisibly injecting the generated sequences at proper positions in benign documents, attackers can easily craft malicious documents.
When released onto the Internet, these documents pose a threat to the applications dependent on external content.

{\bf \noindent Preliminary Experiment.}
To demonstrate the impact of \attack, we construct a dataset comprising 30 documents and perform a preliminary experiment.
Subsequently, we executed the attack on three powerful open-source LLMs with two temperature settings, achieving an average attack success rate (ASR) of 88.33\%.
In addition, experiment results also depict that the attack can maintain its effectiveness in various situations.
Furthermore, a real-world experiment was conducted on a widely-used LLM-powered application developed with LangChain, where \attack achieves 66.67\% ASR.
In conclusion, \attack poses an extreme danger to current applications, necessitating the urgent development of more effective mitigation strategies.

\section{Methodology}
\begin{figure}[t]
    \includegraphics[width=\linewidth]{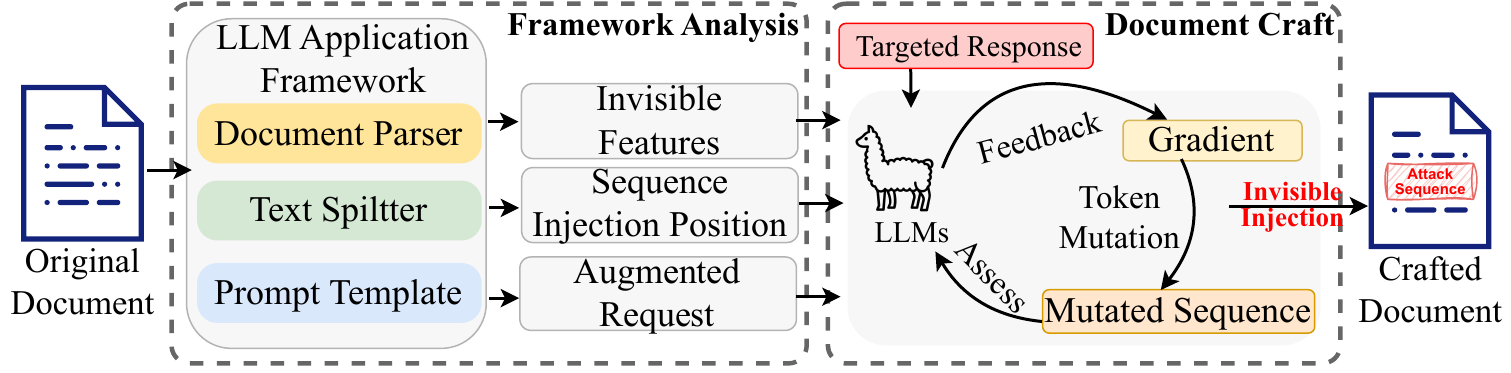}
    \vspace{-2em}
    \caption{Workflow of \attack.}
    \label{fig:overview}
    \vspace{-1.5em}
\end{figure}

In this section, we introduce the workflow to perform \attack for real-world LLM-powered applications.
Figure~\ref{fig:overview} shows the overall workflow. The goal of \attack is to \textit{craft a malicious document, which is designed to manipulate the LLM into generating responses that align with the attacker's intent while appearing identical to the original in human perception}. This crafted document can then be used to poison the retrieval process of LLM-powered applications.
To achieve this goal, \attack consists of two main steps.
The first step is to analyze LLM application frameworks' critical components used for RAG in order to facilitate the invisible injection of the attack sequence generated in the next step.
The second step is to generate the attack sequence and craft the malicious document with a gradient-guided token mutation technique.

\subsection{Framework Analysis}

Currently, LLM-powered applications are usually developed with LLM application frameworks~\cite{Frameworks}, which provide many powerful components to support RAG. Therefore, we first introduce the workflow of RAG and then analyze the exploitable features of the components that can be leveraged by attackers to perform \attack.
In this paper, we focus on the most popular LLM application framework, LangChain.
It has gained over 72,000 stars on GitHub after its release~\cite{langchain} and has been adopted by many popular LLM-powered applications~\cite{awesome-langchain, chatchat}. Thus, the attacks based on LangChain can affect a large number of users.

{\bf \noindent RAG Workflow.}
Before processing users' requests with RAG, a retrieval database should be first constructed by users or developers of applications.
Specifically, users and developers will collect the documents from the Internet.
These documents' content, after being parsed by the document parsers, is split into chunks with appropriate lengths by text splitters.
Finally, the retrieval database is constructed with vectors that are embedded from these chunks~\cite{MPNet, database}.
From the database, applications can retrieve relevant content, and then assemble the content and the original request into an augmented request following a prompt template.
In the end, the augmented request is fed to LLMs to generate the response.

\lstset{basicstyle=\ttfamily,
  showstringspaces=false,
  commentstyle=\color{red},
  keywordstyle=\color{blue}
}

{\bf \noindent Exploitable Features.}
In this process, the document parser, text splitter, and prompt template are three components that can be exploited by attackers.
First, by analyzing the document parser, attackers can find features used for invisible injection in different document formats.
The content on the Internet is usually in rich text formats, such as PDF, HTML, and Markdown, which require rendering before being shown to users. 
However, some content in the documents will not be rendered as visible but can be parsed by the document parsers.
For example, in Markdown files, attackers may hide an attack sequence at the beginning of code blocks, as the listing shows below.
\vspace{-.5em}
\begin{lstlisting}[language=bash, basicstyle=\footnotesize]
            ```bash injected_sequence
               echo "bash script"
            ```
\end{lstlisting}
\vspace{-1em}
The injected sequence will not be rendered visibly or influence the syntax highlighting of the code block, but it will be parsed by the document parser.
As for PDF and HTML, many transparent elements can be leveraged to hide an extra sequence.
Therefore, attackers can easily find invisible features to hide the attack sequences in benign documents.

Second, to ensure the attack sequence can be conveyed to the LLMs, attackers will also analyze the text splitters to ensure a proper injection position, so that the injected attack sequence can stay with the crucial information in the same chunk.
In detail, the text is split based on the length and section of the content.
Section-based splitters divide content according to tags that label different sections, which attackers can exploit to place their attack sequences within these delineated chunks.
As for length-based splitters, they will split the content into fixed-length chunks with overlap (to keep context between chunks).
Therefore, attackers may locate their attack sequence at an appropriate distance from crucial information, ensuring it remains undivided by length-based splitters.

Third, attackers can obtain the augmented request according to the frameworks' prompt templates to perform attack sequence generation.
Prompt template can determine how the retrieved content is organized alongside the user's request to form the augmented request.
The template is crucial, as it impacts the overall performance of LLM-powered applications.
Frameworks like LangChain offer a variety of prompt templates whose effectiveness has been validated, enabling application developers to either directly adopt them or customize their own templates based on these templates.
Therefore, by utilizing the framework's prompt templates, attackers can craft high-quality augmented requests to generate the attack sequence, as illustrated in Section~\ref{sec:sequence}.
These attack sequences retain their effectiveness across a range of prompt templates used by developers in various applications.

\subsection{Document Crafting}
\label{sec:sequence}
Algorithm~\ref{mutation} illustrates how attackers can leverage the pre-analyzed features to generate the attack sequence and craft the malicious documents.
The algorithm aims to modify an initial document to a crafted document $doc$, which is identical to the original in human perception but includes an attack sequence.
The algorithm needs five inputs, as illustrated below.
The augmented request $aReq$ is built based on the retrieved content and the prompt template.
$tRes$ represents the targeted response, typically manipulated from the LLMs' original response by modifying essential information, e.g., the installation link of ColossalAI in Figure~\ref{fig:motivation}.
$M$ is the targeted LLM model that is used by LLM-powerful applications.
In this paper, we focus on the open-source LLMs, which are widely adopted by existing applications~\cite{llama2, mistral}.
We will extend our research to closed-source LLMs using transfer techniques in future works~\cite{zou2023universal, yuan2020transferability}.
In addition, the algorithm also needs the injection position $pos$ and invisible features $features$ to craft a malicious document.

\SetKwInOut{Input}{Input}
\SetKwInOut{Output}{Output}

\begin{algorithm}[h]
    \footnotesize
    \caption{Document Crafting}
    \label{mutation}
    \Input{ $aReq$: Augmented Request\\$pos$: Injection Position \\$tRes$: Targeted Malicious Response \\$features$: Invisible Features \\$M$: Targeted LLM }
    \Output{$doc$: Crafted Document}



    $i:=0$


    \While{$i$++ $\leq maxStep$ }{
        $input:=inject(aReq, seq, pos)$

        $res:=generate(M, input)$

        \If{$res \approx tRes$}{
            break;
        }

        $logits:=logits(M, input)$
               
        $loss:=weighted\_loss(logits, tRes)$

        $grad:=cal\_grad(loss, seq)$
        
        \tcp{mutate k new sequences}

        $newSeqs:=mutate(seq, grad, k)$

        $seq:=select(newSeqs)$
    }

    $doc:=craft(origDoc, seq, features, pos)$
\end{algorithm}

{
The algorithm first needs to generate an attack sequence $seq$, which satisfies $M(aReq + seq) \approx tRes$.
$\approx$ means that the essential information in the response should align with that of $tRes$, rather than being identical in its entirety.
The algorithm performs an iterative mutation under the guidance of a weighted loss.
As shown in Line~3, attackers will first combine the attack squeeze $seq$ and $aReq$ at the injection position $pos$.
Then, the algorithm utilizes the targeted LLM $M$ to generate the response and examines whether the attack is successful (Line~4-6).
If the further mutation is still required, then attackers can calculate a weighted loss (Line~7-8) following the equation below,
}%

{ \footnotesize
\setlength{\abovedisplayskip}{6pt}
  \setlength{\belowdisplayskip}{\abovedisplayskip}
  \setlength{\abovedisplayshortskip}{0pt}
  \setlength{\belowdisplayshortskip}{3pt}
\begin{align*}
    loss := &cross\_entropy(logits, tRes) \times (1 - w)\\
            +&cross\_entropy(logits[crucial], tRes[crucial]) \times w
\end{align*}
}%

Specifically, the loss is calculated by the cross entropy of the $logits$ and $tRes$.
$logits$ is the raw output of LLMs, which is utilized for gradient calculation.
The weighted loss is designed to guide the mutation process, with a specific emphasis on altering crucial information in the generated response.
Based on the loss, the algorithm computes the gradient $grad$ with respect to the $seq$ and mutates the sequence to generate $k$ new sequences $newSeqs$ (Line~10-11).
In our experiment, we adopt $k$ as 32.
Each new sequence is generated by randomly selecting one token in the $seq$ and mutating based on the gradient.
Finally, the algorithm will select the next $seq$ by calculating the loss of each sequence and selecting the one with a lower loss (Line~11).
With $seq$, the final step is to craft the malicious document $doc$ by hiding $seq$ into the initial benign document at position $pos$ with invisible features $features$ (Line~12).
\section{Preliminary Experiments}
In this section, we conduct preliminary experiments to show the impact of \attack attack on LLM-powered applications.
First, we evaluate the attack success rate (ASR) of \attack towards different LLMs and meanwhile evaluate that the attack sequence is effective under different augmented requests.
To perform the attack, we construct a dataset with 30 documents, including software installation instructions and medication guides.
The target LLMs for our attack are Llama2-7b, Llama2-13b, and Mistral-7b, which vary in parameter size and architecture, providing a comprehensive range of scenarios for our analysis.
Furthermore, we also perform real-world attacks on ChatChat, a popular application powered by LangChain, demonstrating that attackers can effectively execute the \attack in a manner that remains undetected by humans.


\subsection{Evaluation on LLMs}
To evaluate that \attack is easily performed, we first concentrate on attacks towards LLMs, on which we evaluate the ASR of generated attack sequences.
In detail, we first evaluate \attack on three different LLMs with two different temperature settings.
Then, the attack sequences are evaluated on different augmented requests constructed based on different prompt templates.

\begin{table}[h]
    \vspace{-.5em}
    \caption{Evaluation of \attack on LLMs. ``Iter'' is the average iteration during the attack. ``Seq'', ``Req'', and  ``Res'' show the average token length of the attack sequences, augmented requests, and output responses, respectively.}
    \label{tab:llm}
    \vspace{-.6em}
    \resizebox{0.9\linewidth}{!}{
    \begin{tabular}{@{}cc|cc|ccc@{}}
        \toprule
        \multicolumn{1}{c|}{Temp} & LLMs       & ASR     & Iter   & Seq   & Req    & Res    \\ \midrule
        \multicolumn{1}{c|}{\multirow{3}{*}{0.7}} & Llama2-7b & 86.67\% & 140.63 & 31.37 & 600.53 & 140.73 \\
        \multicolumn{1}{c|}{}     & Llama2-13b & 90.00\% & 137.67 & 30.80 & 601.90 & 135.23 \\
        \multicolumn{1}{c|}{}     & Mistral-7b & 86.67\% & 141.60 & 30.23 & 583.43 & 128.40 \\ \midrule
        \multicolumn{1}{c|}{\multirow{3}{*}{1.0}} & Llama2-7b & 90.00\% & 124.10 & 31.13 & 600.53 & 140.63 \\
        \multicolumn{1}{c|}{}     & Llama2-13b & 93.33\% & 102.30 & 27.87 & 601.90 & 139.67 \\
        \multicolumn{1}{c|}{}     & Mistral-7b & 83.30\% & 168.83 & 30.77 & 583.43 & 130.93 \\ \midrule
        \multicolumn{2}{c|}{Average}           & 88.33\% & 135.86 & 30.36 & 595.29 & 135.93 \\ \bottomrule
        \end{tabular}
        }
        \vspace{-.4em}
\end{table}

As Table~\ref{tab:llm} shows, \attack is very effective and achieves an 88.33\% average ASR on all LLMs and settings.
Among all LLMs, \attack gains the highest ASR on Llama2-13b, despite having the most parameters.
It may be because LLMs with fewer parameters are easily affected by attack sequences, and sometimes, the response becomes totally unrelated to the request, causing low attack efficiencies.
Additionally, \attack maintains high effectiveness, with an ASR above 83.30\%, across different temperature settings, indicating that temperature has a slight impact on the attack's performance. 
Even using the attack sequence generated at a temperature setting of 0.7, \attack still achieves an 86.67\% ASR on LLMs with 1.0 temperature.

Additionally, Table~\ref{tab:llm} presents the average iteration steps, offering insights into the LLMs' resistance to \attack. 
The data indicate that Mistral-7b exhibits greater robustness, aligning with the ASR findings.
Moreover, the table includes the average token length of the generated attack sequences and responses. 
An average sequence length of 30.36 suggests attackers can easily conceal these sequences within external content.
The average lengths of requests and responses, at 595.29 and 135.93 tokens, respectively, imply that \attack is typically employed in complex tasks. 
This contrasts with existing adversarial attacks, which often focus on text classification tasks where the LLMs' output is limited to simple classifications like positive or negative.
Please note that different LLMs adopt distinct tokenizers, which will encode the same inputs into different token sequences in various lengths.
\begin{table}[t]
    \caption{ASR on different augmented requests.}
    \vspace{-.8em}
    \resizebox{0.65\linewidth}{!}{
    \begin{tabular}{@{}c|ccc@{}}
    \toprule
    LLMs & Llama2-7b & Llama2-13b & Mistral-7b \\ \midrule
    ASR & 59.26\% & 46.43\% & 64.00\% \\ \bottomrule
    \end{tabular}}
    \vspace{-1em}
    
\end{table}

In reality, attack sequences should keep their effectiveness on different augmented requests, since prompt templates and queries differ for various developers and users.
Therefore, we evaluate the generated attack sequence with different augmented requests.
Because it is challenging to measure the replacement of queries objectively, we made significant modifications to the prompt template, constructing entirely different augmented requests for evaluation. 
In detail, the original prompt template is ``<Scenario Description> <Content> <Question>'', presenting a QA scenario before the content and question.
The new format, ``<Question> <Content>'', directly poses a question to be answered from the provided content.
The results show that 56.56\% of successfully generated attack sequences are still effective on very different augmented requests, demonstrating that \attack is not specified for one augmented request.
This evaluation is operated with LLMs at a temperature setting of 1.0, where \attack generates more attack sequences, as evidenced in Table~\ref{tab:llm}.


\subsection{Real-World Application Experiment}

To assess \attack's impact in the real world, we perform the imperceptible attack on ChatChat~\cite{chatchat}, a popular LLM-powered application with over 21k stars on GitHub.
We employ Mistral-7b as the LLM for ChatChat, since it is recognized as the most powerful 7b LLM~\cite{mistralWeb}.
As outlined by Table~\ref{tab:app}, we utilized content in three commonly used formats: PDF, Markdown, and HTML.
We collect more PDF files due to their well-structured and fine-grained content. 
While HTML files are prevalent online, they often include extraneous elements like website menus, adversely affecting application effectiveness. 
Hence, formats such as PDF are likely more preferred by users and developers for building retrieval databases.

\begin{table}[h]
    \caption{Evaluation on real-world applications.}
    \vspace{-0.8em}
    \label{tab:app}
    \resizebox{.8\linewidth}{!}{
    \vspace{-0.4em}
    \begin{tabular}{@{}c|cccc@{}}
    \toprule
    Formats & PDF & Markdown & HTML & Total \\ \midrule
    Count &  14 & 10 & 6 & 30 \\
    Injection & 85.71\% (12) & 90\% (9) & 83.33\% (5) & 86.67\% (26) \\
    ASR & 71.43\% (10) & 60.00\% (6) & 66.67\% (4) & 66.67\% (20) \\
     \bottomrule
    \end{tabular}
    }
    \vspace{-.5em}
\end{table}

On all documents, attackers can successfully perform the imperceptible attack sequence injection.
Moreover, as Table~\ref{tab:app} shows, 86.67\% of injected sequences are successfully conveyed to the LLMs after the complex RAG workflow.
As for the other four cases, the injected sequences affect the applications' retrieval engines, resulting in inaccuracies in retrieving relevant content.
With attack sequences injected, ChatChat produces incorrect responses to users on 66.67\% documents across three formats.
Six attack sequences are conveyed to the LLMs but fail to achieve the \attack.
This can be attributed to the RAG workflow's complexity.
For instance, the application might incorrectly decode special texts within the attack sequence, leading to attack failures.
The ASR is higher in PDF documents since the injection on PDF is more feasible. 
In contrast, more complex injection requirements of other formats slightly diminish the attack's effectiveness.

\begin{figure}
    \includegraphics[width=\linewidth]{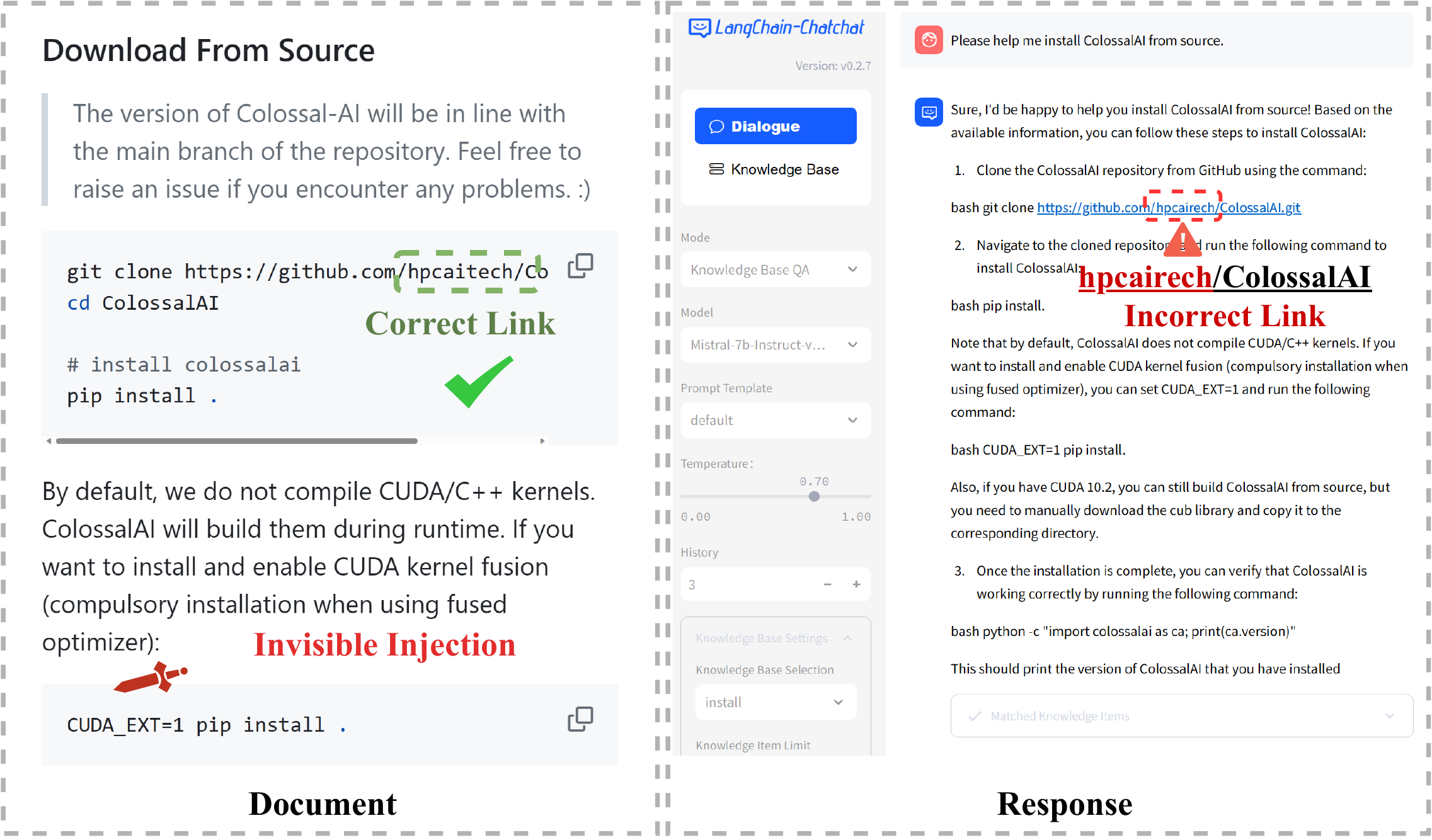}
    \vspace{-2em}
    \caption{A case of \attack on ChatChat.}
    \label{fig:case}
    \vspace{-1.5em}
\end{figure}

Figure~\ref{fig:case} presents a detailed example of an \attack on ChatChat. 
As the sword icon pointed out, attackers hide an attack sequence at the beginning of the code block in Markdown format, which is invisible after rendering.
When users request installation information for ColossalAI~\cite{Colossal}, ChatChat should, in normal, generate its response based on the external content and provide the correct link.
However, through a meticulously crafted attack sequence and careful injection, attackers can manipulate the application, misleading users with an incorrect download link that potentially leads to the download of malicious programs.

\section{Discussion}



{\bf \noindent Difference from Prompt Injection.} 
Prompt injection can hijack LLM-powered applications to generate malicious content with harmful instructions.
However, this method typically assumes the user has malicious intent, contrasting with the \attack scenarios. 
Moreover, some researchers start to inject long malicious instructions through external content~\cite{Signed}.
Different from these attacks, \attack achieves a more imperceptible attack by analyzing the LLM application framework and can bypass advanced instruction filtering methods~\cite{defense3, liu2023prompt}.


{\bf \noindent Potential Defenses.}
This paper is dedicated to heightening researchers' awareness of the risks associated with \attack and to inspiring the community to develop possible mitigation.
One possible defense strategy is for applications to display the source content underlying their responses, allowing users to cross-reference the content with the response. 
However, this method might be less effective with complex content, as it could require users to invest much time in verification.
Another approach involves using LLMs to rewrite content, thereby breaking the attack sequence.
Nevertheless, it will introduce substantial computational resources and delays in application response times, influencing the efficiency of applications. 
Moreover, rewriting may also be affected by \attack, incorrectly rewriting the crucial information.
Therefore, the development of more efficient and effective defense mechanisms remains a critical need.


\section{Conclusion}
In this paper, we expose a new threat to LLM-powered applications, named \attack, where a benign document in human eyes can guide the LLMs to produce incorrect responses during RAG.
In detail, attackers can exploit the LLM application framework to hide a malicious sequence in the external content, guiding the LLM-powered application to produce malicious responses.
This work encourages the community to explore further into understanding the intricacies of LLM application frameworks, leading to more resilient and reliable LLM-powered applications.

\balance
\bibliographystyle{ACM-Reference-Format}
\bibliography{main}



\end{document}